\documentclass[preprint,proceedings]{rmaa}

\SetYear{2005}
\SetConfTitle{RevMexAA, Serie de Conferencias}

\title{Extending the Search for Counterrotating Gas and Stars in Galaxies: A Study of Late-Type Dwarfs}

\author{
  J. M. Guie,\altaffilmark{1} 
  S. J. Kannappan,\altaffilmark{2}
  Z. Balog,\altaffilmark{3}
  and P. Berlind\altaffilmark{4}}

\altaffiltext{1}{University of Texas at Austin, 2511 Speedway, RLM 15.308, Austin, TX 78712, USA (\protect\email{jocelly@mail.utexas.edu}).}
\altaffiltext{2}{University of Texas at Austin, 2511 Speedway, RLM 15.308, Austin, TX 78712, USA (\protect\email{sheila@astro.as.utexas.edu}).}
\altaffiltext{3}{Department of Optics and Quantum Electronics, University of Szeged, Dom ter 9, Szeged, H-6720 Hungary (\protect\email{balogz@titan.physx.u-szeged.hu}).}
\altaffiltext{4}{Fred Lawrence Whipple Observatory, PO Box 97, Amado AZ, 85645, USA (\protect\email{pberlind@cfa.harvard.edu}).}
\suppressfulladdresses

\shortauthor{Guie, Kannappan, Balog \& Berlind}

\listofauthors{J. M. Guie, S. J. Kannappan, Z. Balog \& P. Berlind}
\indexauthor{Guie, J. M.}
\indexauthor{Kannappan, S. J.}
\indexauthor{Balog, Z.}
\indexauthor{Berlind, P.}

\addkeyword{Galaxies: Evolution}
\addkeyword{Galaxies: Kinematics and dynamics}

\begin{document}
\maketitle 

\boldabstract{In a previous study on gas-stellar counterrotation for a large sample of E/S0 and spiral galaxies (Kannappan \& Fabricant 2001), two dwarf irregular galaxies were included, of which one was tentatively identified as a counterrotator. Here we extend the search for counterrotation to include 10 more irregular/spiral dwarf galaxies. We find that all systems with well defined gas and stellar rotation show kinematics consistent with co-rotation. However, we see evidence of decoupled gas and stellar kinematics in $ \ga $ 50\% of the sample, possibly reflecting minor interactions too small to create large-scale gas-stellar counterrotation.}


Except for one dwarf Im, all of the gas-stellar counterrotators found by Kannappan \& Fabricant (2001) were low-luminosity E/S0 systems (Fig. 1). This morphological tendency is consistent with a scenario in which retrograde mergers and interactions between galaxies produce counterrotating gas and stars, because such events also reshape galaxies into more concentrated/centrally bulging E/S0 types. However, no conclusions on counterrotation frequency could be reached for low-luminosity galaxies with irregular/spiral morphologies, due to limited stellar kinematic data.  

Kannappan \& Fabricant's tentative identification of a dwarf irregular counterrotator suggests that luminosity instead of morphology could be the primary correlate with counterrotation, perhaps supporting a scenario in which retrograde gas arrives via efficient ``cold-mode'' gas accretion, which occurs mainly in low mass galaxies (Birnboim \& Dekel 2003). To test these scenarios, we present new stellar kinematic data for 10 late-type dwarf/irregular galaxies, obtained with the FAST spectrograph on the 60'' telescope at Mt. Hopkins, Arizona. Our sample consists of galaxies with B-band effective surface brightnesses between 20.8 and 23.6 (somewhat high for dwarfs). Of these, seven are type Sdm-Im and three are type Pec. Our galaxies reside in global environments ranging from below the mean density of the field, up to nearly the density of the Virgo Cluster, with a median of $ \sim 2\times $ the mean field density (as estimated using code from Grogin \& Geller 1998). We measure gas and stellar kinematics using emission and absorption lines (Fig. 2) in the 4000--6000\,\AA\ range. The gas rotation curves are extracted by simultaneous fitting of the H$\beta$ and nearby [O{\sc III}] emission lines. The stellar rotation curves are obtained by cross correlating with stellar templates spanning a range of metallicities. We homogenize velocity zero points between templates by adopting corrections (0-24 km/s) chosen to minimize gas-stellar velocity differences for the majority of galaxies.

\begin{figure}[!t]
  \includegraphics[width=\columnwidth,height=5 cm]{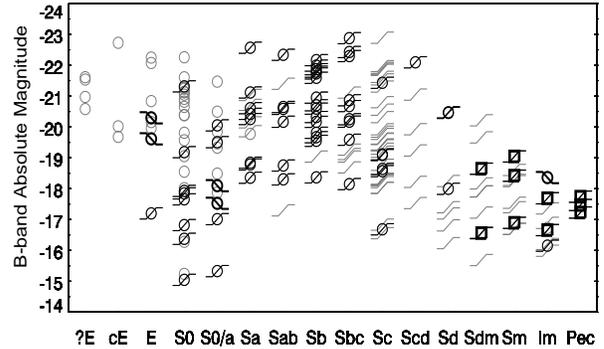}
  \caption{Morphology and luminosity distribution for the Nearby Field Galaxy Survey (Jansen et~al.\ 2001), showing the galaxies with gas and stellar kinematic data from Kannappan \& Fabricant (bent lines and circles, respectively) and with new data from the present work (boxes). Counterrotators are shown with backward bent lines.}
\end{figure}

\begin{figure*}[!t]
  \includegraphics[width=\textwidth,height=11 cm]{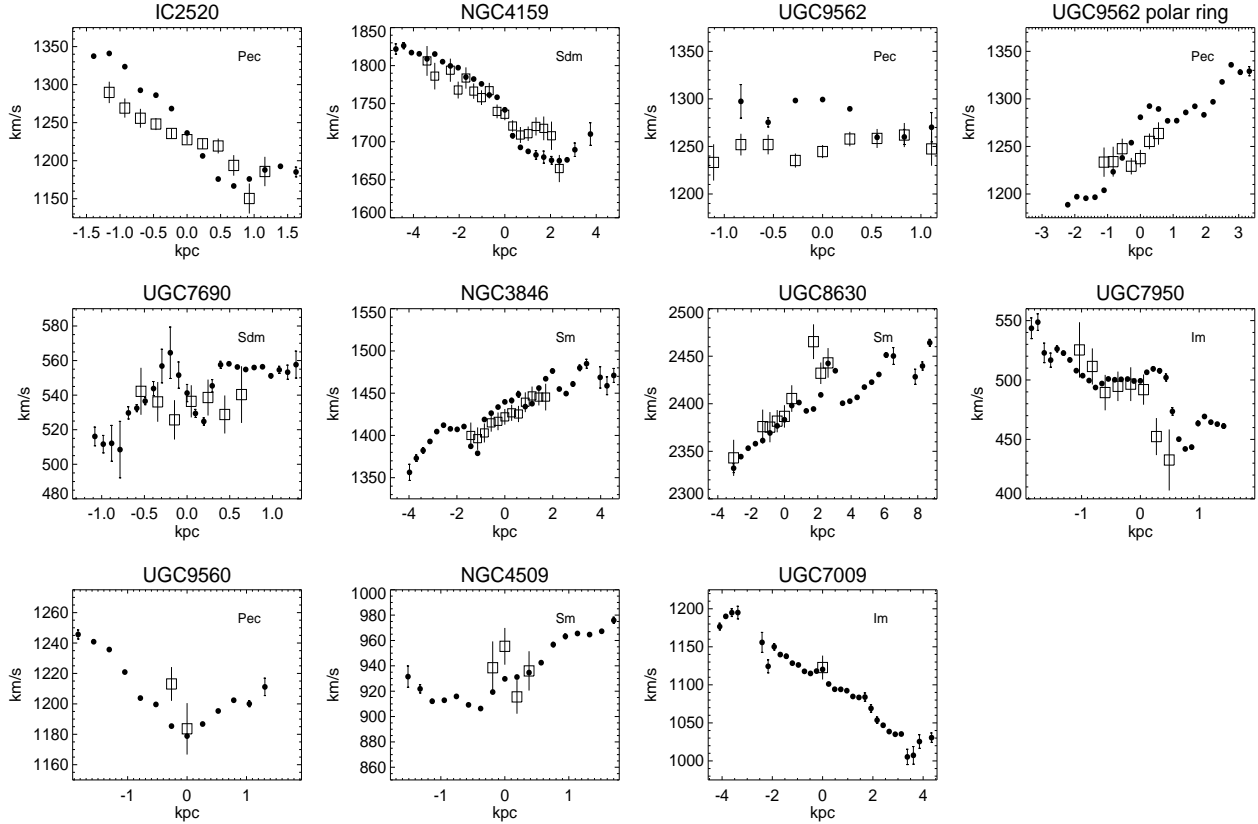}%
   \caption{Gas (dots) and stellar (boxes) rotation curves for our sample, uncorrected for inclination. The stars in IC2520 rotate slower than the gas, suggesting non-coplanar gas and stars and/or late accretion of gas over a hotter stellar component.  UGC9562 is a polar ring galaxy whose gas and stars have systemic velocities offset by $\sim$50 km/s in the middle of the ring.  Cox et al (2001) report that this galaxy is interacting. UGC7690 and NGC3846 show $\sim$20 km/s systemic velocity offsets between their stars and a portion of their gas. Note the two gas components in NGC3846. NGC4159 may show similar decoupling or possibly slow stellar rotation. UGC8630 has two separate stellar continuum peaks with corresponding gas, suggesting merging clumps. UGC7950 also shows kinematic oddities that may reflect decoupling. The stellar kinematics for UGC9560, NGC4509 and UGC7009 are too poorly sampled to draw any conclusions.}
\end{figure*}

All sample galaxies with well defined gas and stellar rotation show kinematics consistent with co-rotation. If the cause of counterrotation is luminosity dependent (efficient gas accretion in low-mass galaxies), then the low frequency of counterrotators in this sample is mildly surprising, but the sample is small. Also, retrograde gas may not survive in such gas-rich systems. If the cause of counterrotation is morphology dependent (mergers/interactions) then these results are as expected. 

Despite the lack of counterrotation, we do see evidence of kinematically decoupled gas and stars in at least five of the late-type dwarfs. The high frequency ($ \ga $ 50\%) of gas-stellar decoupling in our sample could reflect minor interactions too small to create large-scale gas-stellar counterrotation. Of the five galaxies with clear gas-stellar decoupling, three (UGC9562, UGC7690 and NGC3846) have close companions of either comparable or fainter magnitude, all within 70 kpc.\footnote{The two Im galaxies UGC7950 and UGC7009 also have companions within 100 kpc, one brighter and one fainter.} The other two galaxies with gas-stellar decoupling (IC2520 and UGC8630) seem to be merger remnants, one showing merging kinematic sub-components. We note that gas-stellar velocity differences may also be produced by energetic internal disturbances (e.g. supernovae) or by cold-mode accretion arriving in dynamically significant discrete clumps.


\acknowledgements

We are grateful for the help and guidance of Frank Bash and Bev Wills during JMG's undergraduate research experience. We thank Linda Sparke and Avishai Dekel for helpful discussions. JMG was supported by NSF grant AST-0444634. SJK acknowledges an NSF Astronomy \& Astrophysics Postdoctoral Fellowship under grant AST-0401547.

\end{document}